\documentclass[reprint, amsmath, amssymb, aps, prl]{revtex4-2}

\usepackage{graphicx}
\usepackage{pgfplots}
\pgfplotsset{compat=1.16,
    /pgf/declare function={
        lw = 1;
    },
}
\usepgfplotslibrary{groupplots}
\usetikzlibrary{backgrounds}
\usepgfplotslibrary{external}
\tikzsetexternalprefix{tikzext/}

\ifx\runtikz\undefined
    \def\runtikz{0}
\fi

\begin{document}

%\title{Amplitude equations for stored spatially heterogeneous states}
%\title{Amplitude equations of spatially distributed associative memory states from short-range non-local interactions}
\title{Amplitude equations of associative memory patterns in spatially distributed systems}

\author{Akke Mats Houben}
\email{akkemats.houben@ub.edu}
\email{akke@akkehouben.net}
\affiliation{Departament de F\'{i}sica de la Mat\`{e}ria Condensada, Universitat de Barcelona, 08028 Barcelona, Spain}
\affiliation{Universitat de Barcelona Institute of Complex Systems (UBICS), 08028 Barcelona, Spain}

\begin{abstract}
    Evolution equations are derived for the amplitudes of associative memories: heterogeneous states stored in the connectivity of distributed systems with non-local interactions.
    The resulting coupled amplitude equations describe the spatio-temporal dynamics of memory recall.
    They capture pattern completion and selection, and show that short-range connections can sustain spatio-temporal memory pattern dynamics in the form of propagating patterning fronts. 
   The derived amplitude equations are of the same form as those describing classical pattern-forming instabilities, indicating a universality of the dynamics of memory recall and pattern formation in non-equilibrium systems.
\end{abstract}

\maketitle

Many physical and biological phenomena involve the formation and maintenance of spatially patterned states~\cite{haken1977synergetics, cross1993pattern, hoyle2006pattern, cross2009pattern}.
The dynamics of inhomogeneous states arising from isotropic interactions can be understood by studying the linear stability of the homogeneous base-state of the system~\cite{turing1952chemical}, and by performing small-amplitude perturbation analyses around this base-state, leading to amplitude equations capturing the pattern envelope dynamics~\cite{newell1969finite, segel1969distant}.
These show that the possible spatially patterned states that can form are largely determined by the form of the connectivity between the constituent parts, and that pattern evolution is governed by the internal dynamics of these constituents, the specifics of their coupling, and the overlap between the possible inhomogeneous states~\cite{hoyle2006pattern, cross2009pattern}.

Other processes, however, involve the formation of specific spatially heterogeneous states, with the spatial patterns stored in the heterogeneous connectivity between the constituent parts: the recall of stored distributed memories.
Memory storage and recall are commonly modelled using recurrently coupled neural associative memory networks~\cite{little1974existence, hopfield1982neural, peretto1984collective, hopfield1984neurons}, of which modern iterations~\cite{krotov2023anewfrontier} drastically increase memory capacity~\cite{krotov2016dense, demircigil2017onamodel, lucibello2024exponential, vinci2025beyond} and address biological validity~\cite{schonsberg2021efficiency, krotov2021large, podlaski2025storing}.

In this letter amplitude equation derivation is generalised from spatially periodic modes to arbitrary spatially patterned states (memories) stored in the non-local connectivity of distributed systems.
This suggests that memory recall and conventional pattern-forming instabilities belong to a common universality class, and hence formalises a conjectured~\cite{haken1979pattern, baird1986nonlinear} link between associative memory networks, and the theory of pattern formation and dynamics.

The derivation naturally accommodates short-range interactions, making the presented approach well-suited to study memory recall in systems with spatial constraints, such as biological neuronal networks~\cite{braitenberg1991anatomy, perin2011synaptic, campagnola2022local}.
This aligns with previous results showing that associative memories can be stored using connections that decay with distance in some metric space~\cite{roudi2004associative, koroutchev2006bump, roudi2006localized, agliari2015retrieval}.
In addition to these results, the derived amplitude equations allow to study the emerging spatio-temporal memory recall dynamics.

In the following, amplitude equations will be derived for memory patterns in a system extending over a space $\Omega$ for which the activity variable~\footnote{For simplicity of presentation the derivation is presented for a scalar field, but the results are readily extended to multi-dimensional dynamics.} $u(\mathbf{r},t)$ at a location $\mathbf{r} \in \Omega$ evolves under the influence of local internal dynamics $f(u)$ and non-local recurrent coupling,
\begin{equation}\label{eq:du}
    \frac{\partial u(\mathbf{r}, t)}{\partial t} = f(u) + \gamma \int_{\Omega} J(\mathbf{r}, \mathbf{r}') g[u(\mathbf{r}', t)] d\mathbf{r}'.
\end{equation}
The non-local (but possibly spatially constrained by a non-negative spatial kernel $\Phi$) connectivity 
\begin{equation}
    J(\mathbf{r}, \mathbf{r}') = \Phi(|\mathbf{r}-\mathbf{r}'|) \sum_{k=1}^{M} \frac{\mu_k(\mathbf{r}) \mu_k(\mathbf{r}')}{M},
\end{equation}
with $\lVert \mu_k \rVert = 1$, $\int_{\Omega} \mu_k(\mathbf{r})d\mathbf{r} = 0$, and $\int_{\Omega} \Phi(\mathbf{r}) d\mathbf{r} = 1$, stores $M$ spatially heterogeneous states $u \propto \mu_k$ that the system can attain, and it will be for these states that the amplitude equations describe their evolution dynamics.
Nearly orthogonal memory patterns will be assumed.
The parameter $\gamma$ determines the global coupling strength.

Assuming that the system in Eq.~(\ref{eq:du}) has a base-state $u(\mathbf{r},t)=u_0$ unstable to perturbations of the form $u = u_0 + \varepsilon A_k \mu_k$ above some critical coupling~\footnote{due to the normalisation of the memory patterns, multiple $\gamma_k$ typically lie within a small distance from each other.}
\begin{equation*}
    \gamma > \gamma_k = a_1 \left[ b_1 \left\langle\int_{\Omega} J(\mathbf{r}, \mathbf{r}') \mu_k(\mathbf{r}') d\mathbf{r}', \mu_k \right\rangle \right]^{-1},
    %\gamma > \gamma_k = a_1 \left[ b_1 \int_{\Omega}\int_{\Omega} J(\mathbf{r}, \mathbf{r}') \mu_k(\mathbf{r}') d\mathbf{r}' \mu_k(\mathbf{r}) d\mathbf{r}\right]^{-1},
\end{equation*}
with $a_1$ and $b_1$ being the first derivative of, respectively, $f(u)$ and $g(u)$ with respect to $u$ evaluated at $u=u_0$, 
then for coupling close to the critical coupling $\gamma = (1+\varepsilon^2)\gamma_k$, with $\varepsilon^2 \ll 1$, the system dynamics on slow temporal $\tilde{t} = \varepsilon^2 t$ and long spatial $\tilde{\mathbf{r}} = \varepsilon \mathbf{r}$ scales can be captured by the pattern amplitudes $A_k(\tilde{\mathbf{r}}, \tilde{t})$.

\emph{Amplitude equations.\textemdash}%
The derivation of the amplitude equations is analogous to that for spatially periodic patterned states (see~\cite{hoyle2006pattern, cross2009pattern}).
For clarity of exposition it will be assumed that $u_0=0$, and that $f(u)$ and $g(u)$ are odd functions around $u_0$ so that their even-order derivatives vanish at the base-state $u_0$. %~\footnote{The derivation is still valid for more general $u_0$, $f(u)$ and $g(u)$. In this case the perturbation analysis possibly has to be carried out to higher order since it can result in subcritical amplitude dynamics.}.
Introducing the slow spatial $\tilde{\mathbf{r}}$ and temporal $\tilde{t}$ scales, and the perturbation expansion $U \sim u_0 + \varepsilon U_1 + \varepsilon^2 U_2 + \ldots$, with $\varepsilon U_1 = \sum_{k=1}^{M} A_k(\tilde{\mathbf{r}}, \tilde{t}) \mu_k(\mathbf{r})$, will give a system of $M$ coupled evolution equations for the pattern amplitudes $A_k$ as solvability conditions for the $O(\varepsilon^3)$ terms as follows.

Writing $u(\mathbf{r}, t) = U\left(\mathbf{r}, \tilde{\mathbf{r}}, \tilde{t}\right)$, Eq.~(\ref{eq:du}) becomes
\begin{equation*}
    \varepsilon^2 \frac{\partial U}{\partial \tilde{t}} = f(U) + (1+\varepsilon^2)\gamma_k \int_{\Omega} J(\mathbf{r}, \mathbf{r}') g[U(\mathbf{r}', \tilde{\mathbf{r}}', \tilde{t})] d\mathbf{r}'.
\end{equation*}
Then, plugging in the perturbation expansion, expanding $f(U)$ and $g(U)$ around $U=u_0$,
and expanding $U_i$ terms under the integral at $\tilde{\mathbf{r}}' = \tilde{\mathbf{r}}$, gives at $O(\varepsilon)$ the linearised dynamics
\begin{equation*}
    \frac{\partial U_1}{\partial \tilde{t}} = a_1 U_1 + \gamma_k b_1 \int_{\Omega} J(\mathbf{r}, \mathbf{r}') U_1(\mathbf{r}', \tilde{\mathbf{r}}, \tilde{t}) d\mathbf{r}',
\end{equation*}
and at $O(\varepsilon^3)$ the solvability criterion~\footnote{Given that $\Phi$ is symmetric and by assuming memory patterns with sufficient spatial variation on average so that $J(\mathbf{r}, \mathbf{r}')$ will be symmetric around $\mathbf{r}'=\mathbf{r}$, terms involving odd-order derivatives of $U_1$ in the expansion vanish under the integral.}
%\begin{widetext}
%\begin{equation*}
%    \frac{\partial U_1}{\partial \tilde{t}} = \gamma_k b_1 \int_{\Omega} J(\mathbf{r}, \mathbf{r}') U_1(\mathbf{r}', \tilde{\mathbf{r}}, \tilde{t}) d\mathbf{r}' + a_3 U_1^3 + \gamma_k b_3 \int_{\Omega} J(\mathbf{r}, \mathbf{r}') \left[U_1(\mathbf{r}', \tilde{\mathbf{r}}, \tilde{t})\right]^3 d\mathbf{r}' + \gamma_k b_1 \sum_{i=1}^{\dim{\Omega}} \int_{\Omega} J(\mathbf{r}, \mathbf{r}') \frac{(r_i'-r_i)^2}{2} d\mathbf{r}'~ \frac{\partial^2 U_1}{\partial \tilde{r}_i^2},
%\end{equation*}
%\end{widetext}
\begin{equation*}
    \begin{split}
    \frac{\partial U_1}{\partial \tilde{t}} =&~ \gamma_k b_1 \int_{\Omega} J(\mathbf{r}, \mathbf{r}') U_1(\mathbf{r}', \tilde{\mathbf{r}}, \tilde{t}) d\mathbf{r}'  
        + a_3 U_1^3  \\
        &+ \gamma_k b_3 \int_{\Omega} J(\mathbf{r}, \mathbf{r}') \left[U_1(\mathbf{r}', \tilde{\mathbf{r}}, \tilde{t})\right]^3 d\mathbf{r}' \\
        &+ \gamma_k b_1 \sum_{i=1}^{\dim{\Omega}} \int_{\Omega} J(\mathbf{r}, \mathbf{r}') \frac{(r_i'-r_i)^2}{2} \frac{\partial^2 U_1}{\partial \tilde{r}_i^2} d\mathbf{r}',
    \end{split}
\end{equation*}
where $a_3 = f'''(u_0)/6$ and $b_3 = g'''(u_0)/6$.
Finally, multiplying the $O(\varepsilon^3)$ evolution equation of $U_1$ by $\mu_k$ and integrating over the domain gives a system of coupled amplitude equations (in original variables):
%\begin{widetext}
%\begin{equation}\label{eq:ampleq}
%    \frac{\partial A_k}{\partial t} = \varepsilon^2\sum_{m=1}^{M} \alpha_{km} A_m
%    + \sum_{i=1}^{M} \sum_{j=1}^{M} \sum_{m=1}^{M} \beta_{kijm} A_i A_j A_m 
%    + \sum_{m=1}^{M} \frac{D_{km}}{2} \nabla^2 A_m
%\end{equation}
%\end{widetext}
\begin{equation}\label{eq:ampleq}
    \begin{split}
    \frac{\partial A_k}{\partial t} = \varepsilon^2\sum_{m=1}^{M} \alpha_{km} A_m 
    & + \sum_{i=1}^{M} \sum_{j=1}^{M} \sum_{m=1}^{M} \beta_{kijm} A_i A_j A_m  \\
    &+ \sum_{m=1}^{M} \frac{D_{km}}{2} \nabla^2 A_m
    \end{split}
\end{equation}
with coefficients
\begin{equation*}
    \alpha_{km} = \gamma_k b_1 \left\langle \int_{\Omega} J(\mathbf{r}, \mathbf{r}') \mu_m(\mathbf{r}') d\mathbf{r}', \mu_k \right\rangle,
\end{equation*}
\begin{equation*}
    \beta_{kijm} = \left\langle \int_{\Omega} \left[a_3 + \gamma_k b_3 J(\mathbf{r}, \mathbf{r}')\right] \mu_i(\mathbf{r}') \mu_j(\mathbf{r}') \mu_m(\mathbf{r}') d\mathbf{r}', \mu_k \right\rangle,
\end{equation*}
%and the $i$-th element of the vector $\mathbf{D}_{km}$,
%\begin{equation*}
%    D_{km}^{(i)} = \gamma_k b_1 \left\langle \int_{\Omega} J(\mathbf{r}, \mathbf{r}') (r'_i - r_i)^2 \mu_m(\mathbf{r}') d\mathbf{r}', \mu_k \right\rangle.
%\end{equation*}
and, since $\Phi$ is isotropic and assuming the memories to be nearly isotropic on average leads to isotropic diffusion, 
\begin{equation*}
    D_{km} = \gamma_k b_1 \left\langle \int_{\Omega} J(\mathbf{r}, \mathbf{r}') (\mathbf{r}'-\mathbf{r})^2 \mu_m(\mathbf{r}') d\mathbf{r}', \mu_k \right\rangle.
\end{equation*}

\emph{Pattern dynamics.\textemdash}%
The $M$ coupled equations, Eq.~(\ref{eq:ampleq}), describe the growth, interactions and spatio-temporal dynamics of the $M$ stored memory patterns in the full system.
The coupled amplitude equations constitute a gradient system $\partial \mathbf{A}/\partial t = - \nabla V[\mathbf{A}]$, given the potential
\begin{equation*}
    V[\mathbf{A}] = \int_{\Omega} \sum_{k=1}^{M} \left( -F_k[\mathbf{A}] + \frac{1}{2}\sum_{m=1}^{M} \nabla A_k \cdot D_{km} \nabla A_m \right) d\mathbf{r},
\end{equation*}
with
\begin{equation*}
    F_k[\mathbf{A}] = \sum_{m=1}^{M} \left(\frac{\varepsilon^2 \alpha_{km}}{2}
    + \sum_{i=1}^{M} \sum_{j=1}^{M}  \frac{\beta_{kijm}}{4} A_i A_j \right) A_m A_k.
\end{equation*}
Hence, the system at $\varepsilon \mathbf{r}$ spatial and $\varepsilon^2 t$ temporal scales will tend to settle into a time-independent state.

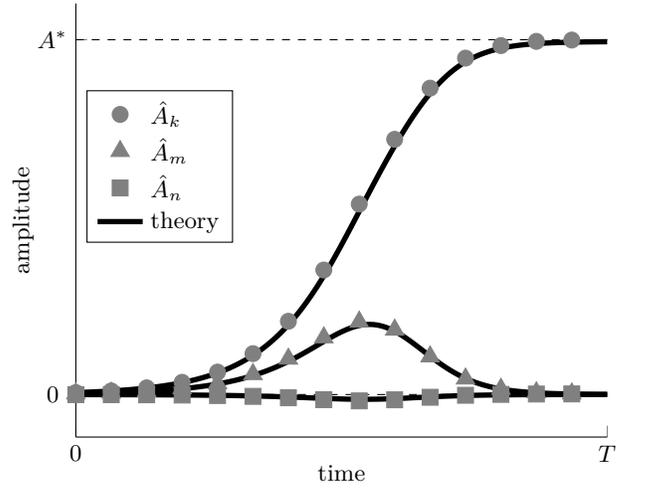
\begin{figure}
    \centering
    \if\runtikz1
        \begin{tikzpicture}
            \begin{axis}[
                    width=1\columnwidth,
                    height=0.85\columnwidth,
                    title={~},
                    xlabel={time},
                    xlabel style={yshift=5pt},
                    xmin=0, xmax=3000,
                    xtick={0, 3000},
                    xticklabels={{$0$}, {$T$}},
                    axis x line*=bottom,
                    ylabel={amplitude},
                    ylabel style={yshift=-5pt},
                    ytick={0, 1.415},
                    yticklabels={$0~$, $A_k^{\infty}$},
                    axis y line*=left,
                    legend entries={$\hat{A}_k$, $\hat{A}_m$, $\hat{A}_n$, theory},
                    legend style={at={(0.02, 0.8)}, anchor=north west},
                    legend cell align={left},
                ]
                \addplot[color=gray, mark=*, only marks, mark repeat=1000, mark size=3] table[x=t, y=u1]{./data/results_bulk_M3_100.dat};
                \addplot[color=gray, mark=triangle*, only marks, mark repeat=1000, mark size=4] table[x=t, y=u2]{./data/results_bulk_M3_100.dat};
                \addplot[color=gray, mark=square*, only marks, mark repeat=1000, mark size=3] table[x=t, y=u3]{./data/results_bulk_M3_100.dat};

                \addplot[color=black, line width=2] table[x=t, y=A1]{./data/results_bulk_M3_100.dat};
                \addplot[color=black, line width=2] table[x=t, y=A2]{./data/results_bulk_M3_100.dat};
                \addplot[color=black, line width=2] table[x=t, y=A3]{./data/results_bulk_M3_100.dat};
                \addplot[black, dashed, domain=0:3000, samples=2] {1.415};
                \addplot[black, dashed, domain=0:3000, samples=2] {0};
            \end{axis}
        \end{tikzpicture}%
    \else
        \includegraphics{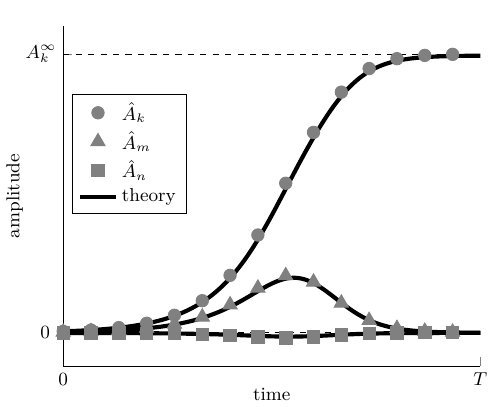}
    \fi
    \caption{Bulk dynamics with $M=3$ stored patterns, comparing the pattern amplitudes $\hat{A}_i = \langle u(t,x), \mu_i\rangle$ obtained from simulating the full system [Eq.~(\ref{eq:du}), grey markers], and the amplitudes obtained by numerically integrating the system of amplitude equations [Eq.~(\ref{eq:ampleq}), black lines].}\label{fig:bulk}
\end{figure}

Bulk dynamics of the memories in the full system (grey markers in Fig.~\ref{fig:bulk}) are captured by the amplitude equations (black lines in Fig.~\ref{fig:bulk}), describing the pattern growth from low amplitude initial conditions, their transient dynamics, and the subsequent saturation of a single pattern to amplitude $A_k \to A_{k}^{\infty} = \varepsilon \sqrt{\alpha_{kk}/\beta_{kkkk}}$ together with the suppression of the other patterns $A_{m \neq k} \to 0$ due to the cubic interaction terms.
For small enough $\varepsilon$ the coefficients of the linear terms $A_k$ and cubic interactions $A_m^2 A_k$ in Eq.~(\ref{eq:ampleq}) will balance, so that in general the only stable equilibria will be single pattern states, as in the example of Fig.~\ref{fig:bulk}.
Symmetric states, with multiple patterns having the same non-vanishing amplitude in a single bulk domain, are unstable equilibria.

\begin{figure}
\centering
    \if\runtikz1
        \begin{tikzpicture}
            \begin{axis}[
                title={~},
                view={0}{90},
                width=0.65\columnwidth,
                height=0.85\columnwidth,
                enlargelimits=false,
                tick style={draw=none},
                xlabel={space},
                xlabel style={yshift=10pt},
                xtick={-157, 157},
                xticklabels={{$-L$}, {$L$}},
                axis x line*=bottom,
                ylabel={time},
                ylabel style={yshift=-2pt},
                ytick={0, 247},
                yticklabels={0, $T$},
                ymin={0}, ymax={247},
                axis y line*=left,
                ylabel style={yshift=-5pt},
                y dir=normal,
                %
                %legend entries={{$u \odot \mu_k$}, theory},
                %legend pos=south east,
                %legend style={fill=gray!50},
                %legend cell align={left},
                %
                colorbar,
                point meta min=0,
                point meta max=1,
                colormap/viridis,
                colorbar style={
                    width=10pt,
                    xshift=-5pt,
                    ylabel={{$\hat{a}_k$}},
                    ylabel style={yshift=15pt},
                    ytick={0, 1},
                    yticklabels={{$0$}, {$A^{\infty}_k$}}
                },
                ]
                %\addplot[plot graphics/node/.append style={yscale=-1, anchor=north west}] graphics[xmin=-94, xmax=94, ymin=0, ymax=100] {figs/ampleq_tw.pdf};
                \addplot[plot graphics/node/.append style={yscale=-1, anchor=north west}] graphics[xmin=-157, xmax=157, ymin=0, ymax=247] {figs/ampleq_tw_2.pdf};
                \addplot[
                    domain=0:247,
                    samples=200,
                    white,
                    dashed,
                    line width=3,
                    parametric,
                ] ({-8.4 - 0.42276744133982774*x}, x);
                \addplot[
                    domain=0:247,
                    samples=200,
                    white,
                    dashed,
                    line width=3,
                    parametric,
                ] ({8.4 + 0.42276744133982774*x}, x);
            \end{axis}
        \end{tikzpicture}
    \else
        \includegraphics{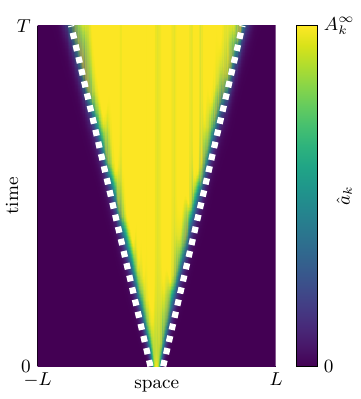}
    \fi
    \caption{Spatio-temporal profile of a pattern forming wave-front invading a spatially homogeneous domain. 
        Colour intensity indicates the local pattern amplitude $\hat{a}_k(r,t) = 2L \left[u(r, t) \odot \mu_k(r)\right]$. 
    The theoretical front positions $r_\pm^{(k)}(t) = r_\pm^{(k)}(0) \pm \varepsilon \sqrt{2 \alpha_{kk} D_{kk}}t$ are shown as white dashed lines.
    }\label{fig:patwave}
\end{figure}
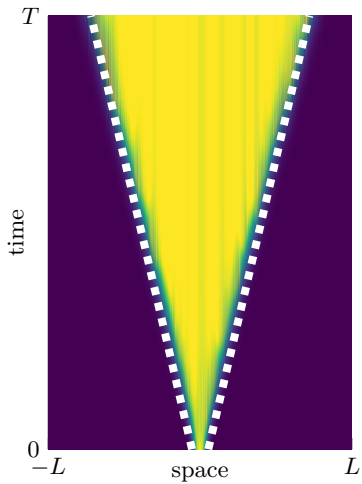

Spatial restriction of the connections ($\Phi$ vanishing for large arguments) leads to a finite diffusion rate of the memory patterns, and hence spatio-temporal pattern dynamics are expected in the full system.
The equations governing the pattern amplitudes [Eq.~(\ref{eq:ampleq})] are $M$ coupled parabolic partial differential equations, so travelling wave-front solutions, connecting $A_k = A_{k}^{\infty}$ behind the wave-front to $A_k = 0$ infront of the wave-front, exist~\cite{fife1979mathematical, volpert1994traveling, saarloos2003front}.
These solutions correspond to a patterned solution, $u(\mathbf{r}, t) \propto \mu_k(\mathbf{r})$, invading an unpatterned spatially homogeneous domain (illustrated with the colour intensity in Fig.~\ref{fig:patwave}), in analogy to patterning fronts for spatially periodic states~\cite{dee1983propagating} and reflecting experimental evidence incidating travelling disturbances are involved in the recall and consolidation of memories in brains~\cite{patel2013local}.
The linear spreading velocity for a plane wave-front can be found by standard methods to be $c_k = \varepsilon \sqrt{2\alpha_{kk} D_{kk}}$.
The theoretical front positions, using $c_k$, are indicated with white dashed lines in Fig.~\ref{fig:patwave} and agree with the fronts in the full system.

Short-range connectivity also allows the contemporaneous retrieval of multiple different patterns in separated domains of the system~\cite{roudi2004associative, koroutchev2006bump, roudi2006localized, agliari2015retrieval}: a state of multiple domains displaying different patterns separated by domain walls corresponds to a stable configuration of the system due to the normalisation of the memory patterns.
In the case of memory patterns stored with differing strengths, spatio-temporal memory selection dynamics can emerge in the form of patterning fronts in which a strong memory invades the domain of a weaker memory.

\emph{Conclusions.\textemdash}%
It is shown that the derivation of pattern amplitude equations for spatially periodic inhomogeneous states extends naturally to arbitrary spatially heterogeneous states that are stored in the (possibly short-range) non-local connectivity of the system.
The resulting amplitude equations retain the same form as those for spatially periodic states, but with mode interactions affecting linear growth (non-vanishing off-diagonal entries in $\alpha_{km}$) and leading to cross-diffusion (non-vanishing off-diagonal entries in $D_{km}$).

The memory pattern dynamics are universal for systems of the form Eq.~(\ref{eq:du}), and are governed by the derived system of coupled amplitude equations [Eq.~(\ref{eq:ampleq})] in case the spatially homogeneous base-state of the system looses stability supercritically.
To derive the subcritical amplitude equations the perturbation expansion needs to be carried out to higher order in order to capture terms that saturate the pattern growth.

The ability of the derived amplitude equations to accurately capture the spatio-temporal dynamics of memory recall underscores a strong connection between pattern formation and associative memory.
This invites the application of other insights from pattern formation theory to study memory storage, recall, and selection in associative memory models and, given the natural handling of short-range connections, to the study memory in physical and biological systems.% and, in particular, in systems with spatial restrictions and short-range connectivity.

\bibliography{library.bib}
\end{document}